\documentclass[12pt,reqno]{article}
\usepackage[lmargin=2cm, rmargin=2cm]{geometry}
\usepackage{amsfonts}
\usepackage{amssymb}
\usepackage{enumerate}
\usepackage{amsthm}
\usepackage{array}
\usepackage{epsfig}
\usepackage[small,bf]{caption}
\usepackage{appendix}
\usepackage{dsfont}
\usepackage{hyperref}
\usepackage{amsmath}
\usepackage{graphicx}
\usepackage{epstopdf}
\usepackage{blkarray}

\allowdisplaybreaks \numberwithin{equation}{section}

\begin{document}

\begin{titlepage}
 \thispagestyle{empty}

\begin{flushright}
    % \hfill{ITP-UU-14/29 }\\
     %  \hfill{CPHT-RR098.1214 }\\
 \end{flushright}

 \begin{center}

 \vspace{30mm}

     { \LARGE{\bf  {Black holes as random particles: entanglement dynamics in infinite range and matrix models}}}

     \vspace{40pt}

\Large{{\bf Javier M. Mag\'an}} \\[8mm]
{\small\slshape
Institute for Theoretical Physics \emph{and} Center for Extreme Matter and Emergent Phenomena, \\
Utrecht University, 3508 TD Utrecht, The Netherlands \\

\vspace{5mm}

{\upshape\ttfamily javier.magan@cab.cnea.gov.ar}\\[3mm]}

\vspace{8mm}

     \vspace{10pt}

    \vspace{10pt}

\date{\today}

\end{center}

\begin{abstract}
We first propose and study a quantum toy model of black hole dynamics. The model is unitary, displays quantum thermalization, and the Hamiltonian couples every oscillator with every other, a feature intended to emulate the color sector physics of large-$\mathcal{N}$ matrix models. Considering out of equilibrium initial states, we analytically compute the time evolution of every correlator of the theory and of the entanglement entropies, allowing a proper discussion of global thermalization/scrambling of information through the entire system. Microscopic non-locality causes factorization of reduced density matrices, and entanglement just depends on the time evolution of occupation densities. In the second part of the article, we show how the gained intuition extends to large-$\mathcal{N}$ matrix models, where we provide a gauge invariant entanglement entropy for `generalized free fields', again depending solely on the quasinormal frequencies. The results challenge the fast scrambling conjecture and point to a natural scenario for the emergence of the so-called brick wall or stretched horizon. Finally, peculiarities of these models in regards to the thermodynamic limit and the information paradox are highlighted.
\end{abstract}

 \vspace{10pt}
\noindent

\end{titlepage}

\thispagestyle{plain}

\tableofcontents

\baselineskip 6 mm

\newpage

\section{Introduction and description of results}\label{secI}

The purpose of this article is twofold. We first want to propose and study a quantum model that captures several important features of black hole dynamics and it is still tractable from a unitary microscopic perspective. The desire is to collect enough and robust intuition through a concrete solvable model. The second objective is to extend this intuition to exact descriptions of black holes, such as large-$\mathcal{N}$ Matrix Models\footnote{In this article we will use $\mathcal{N}$ to denote the size of the matrices in the matrix model. The number of degrees of freedom $N$ will scale as $N\sim \mathcal{N}^{2}$. We choose this uncommon convention so that $N$ measures the entropy both in the toy model, in which it just counts the number of oscillators, and in the matrix model.} \cite{thooft,matrix,adscft}.

The essential features we want our toy model to possess are the following:
\begin{itemize}
\item The model must display quantum thermalization \cite{useth}.
\item The model must have the strongest form of non-locality: the Hamiltonian must couple every oscillator with every other oscillator through couplings of the same size \cite{susskind}. Throughout the article we will call these type of models \emph{democratic}, to distinguish between other possible forms of non-locality \footnote{Non-locality can appear in diverse non-equivalent forms. One famous example is noncommutative geometry and noncommutative quantum field theory \cite{ncom}. Other  examples are systems with long range couplings preserving translational symmetry \cite{simone}.}.
\end{itemize}
The first bullet point is an obvious necessary feature.  Since the groundbreaking contributions  of Bekenstein \cite{bekenstein} and Hawking \cite{hawking}, it is widely accepted that black holes are thermal objects. With the advent of the AdS/CFT correspondence \cite{adscft}, the previous statement has a firm theoretical ground, since AdS/CFT maps black hole physics into a many-body quantum system at finite temperature \cite{witten}.

The second bullet point is less obvious. Within conjectured non-perturbative definitions of string theory \cite{matrix,adscft}, black holes are mapped to high energy states of large-$\mathcal{N}$ matrix models. These matrix models have interaction terms of the following form:
\begin{equation}\label{matrixh}
\sum\limits_{i,j=1}^{\mathcal{N}}\pi_{i}\,A_{ij}\,\pi_{j}\:,
\end{equation}
comprising interactions between fundamental and adjoint (matrix valued) degrees of freedom, see \cite{matrixpol} for an explicit example applied to quantum black holes. In the high energy sector of the model, the matrix-valued field $A_{ij}$ has all its entries thermally excited. From the point of view of the vector field $\pi_{i}$, the previous term connects every oscillator $\pi_{i}$ with every other $\pi_{j}$ democratically. Another argument supporting the use of democratic models comes from the recently proposed microscopic model of certain black hole dynamics \cite{subir}, in which a set of fermions interact through random quartic interactions.

Given the previous set of arguments and requirements, the toy model of black hole dynamics we propose in this article is defined by the following Hamiltonian:
\begin{equation}\label{H}
H=\alpha \sum\limits_{i=1}^{N}c_{i}^{\dagger}\,c_{i}+\eta\sum\limits_{i,j=1}^{N}c_{i}^{\dagger}\,V_{ij}\,c_{j}\;,
\end{equation}
where $\alpha$ and $\eta$ are parameters with dimensions of energy, $c_{i}^{\dagger}$ and $c_{i}$ are creation and annihilation operators of spinless free fermions (with the usual anticommutator relations), and $V_{ij}$ are independent random gaussian real numbers with zero mean and unit variance \footnote{The gaussian nature of each matrix entry can be relaxed. As customary in random matrices, the only thing that matters is the mean and the variance of each entry. What it is important in our approach is that all entries are random with the same variance. Relaxing this condition is an interesting open problem we leave for future work. In particular, regarding \cite{useth}, it is interesting to ask how many couplings can we switch off (or correlate with other couplings) so that ergodicity is broken. For the concerns of this article, the most important feature the model possess is large-N factorization. It would be interesting to see to what extent we can make the model less random but still mantaining large-N factorization. We thank an anonymous referee for pointing this aspect to us.}. Therefore, the matrix $(\eta V)_{ij}\equiv\eta V_{ij}$ is a random matrix taken from the Gaussian Orthogonal Ensemble (GOE) with deviation $\sigma_{\eta V}=\eta$. For a beautiful and modern treatment of random matrices see \cite{tao}. Notice that considering bosons instead of fermions is straightforward. In second-quantized formalism all the computations are similar, and one obtains the same conclusions.

The eigenstate correlations and entanglement patterns of the family of Hamiltonians~(\ref{H}) were studied in \cite{useth}, where it was analytically proved that the model satisfies the so-called Eigenstate Thermalization Hypothesis (ETH) \cite{srednickisub} \footnote{See \cite{ETH} for an earlier discovery of ETH and \cite{eisertreview} for an extensive review of the subject of quantum thermalization.}. This hypothesis concerns the origin of statistical mechanics in closed quantum systems. As remarked in \cite{useth}, the family of Hamiltonians~(\ref{H}) shows that ETH is `typical' within the space of \emph{gaussian} systems. Non-linear interactions are not needed in this regard, contrary to common belief.

Below we will study two different types of non-equilibrium unitary processes which are particularly interesting from the point of view of black hole physics. We first consider the typical black hole thought experiment, in which we `throw in' some small system to a black hole. This amounts to set an initial state in which a small subsystem is factored out from the thermal ensemble~(\ref{initial}). We then compute the time evolution of two-point correlation functions and entanglement entropies~(\ref{S}),~(\ref{dS}),(\ref{cort}), and~(\ref{entt}). The fact that our Hamiltonian is gaussian implies that the computation of all the two-point correlation functions provides every $n$-point correlation function as well. This will allow us to discuss properly any subtle question regarding information spreading. The second type of initial states we consider are completely factorized states~(\ref{factorized}). This non-equilibrium process intends to emulate `collapse' scenarios in black hole physics, in which a set of particles with definite quantum numbers collapse beyond their Schwarzschild's radius. This situation turns out to be severely more complicated. Although we are not able to solve for the evolution of occupation densities exactly, we will show interesting relations that the evolution of entanglement entropies satisfy, see~(\ref{scaling}),~(\ref{scaling1}),~(\ref{fermions}) and~(\ref{bosons}). These special relations were argued to hold generically for democratic systems in \cite{uswalks}.

In section~(\ref{secIII}) we apply the results to black hole physics. We first analyze the toy model. Without taking into account Poincar\'e recurrences, the model displays three types of thermalization time scales. The first one is the local relaxation time scale, the analog of quasinormal decay. It is given by $t_{\textrm{local}}=1/R$, where $R^{2}\equiv 4N\eta^{2}\sim \mathcal{O}(\alpha)$ can be taken to be independent of the system size. The second one is the scrambling time scale \cite{hayden,susskind}, defined equivalently as the time by which entanglement entropies become thermal or the time by which information spreads through the whole system, as it was defined in the original reference \cite{susskind}. It can also be defined as the time at which the mean field approximation ceases to apply. It turns out to be given by $t_{\textrm{scrambling}}=t_{\textrm{local}}=1/R$, a result that challenges the fast scrambling conjecture, as we describe properly below. Finally, we notice the existence of another time scale. It is the time by which deviations from thermality are uniform over the entire system. This `randomization' time scale turns out to be $t_{\textrm{random}}=N^{1/3}/R$ for our toy model.

In the second part of section~(\ref{secIII}), we extend the results and intuition to exact descriptions of black hole dynamics, i.e to large-$\mathcal{N}$ matrix models or the SYK model \cite{subir}. The extension is possible because the results of the toy model are rooted
in a single feature: large-$N$ factorization during time evolution in high energy states. This feature causes extensive entanglement dynamics. For the SYK the extension is trivial. For the matrix model it works in Fourier space, providing a gauge invariant notion of entanglement entropy associated to a given set of `generalized free field' modes \footnote{Generalized free fields are gauge invariant operators that generate a free Fock space in the large-N limit. See section~(\ref{secIII}) below for a brief description and Ref.~\cite{papadodimas} for a detailed treatment.}. If we have a set $A$ of generalized free fields modes $\mathcal{O}_{\omega, k}^{i}$ with associated number operator $n_{\mathcal{O}_{\omega, k}^{i}}$, the entanglement entropy is given by:
\begin{equation}\label{SO}
S_{A}(t)=\sum\limits_{i\in A}(\,n_{\mathcal{O}_{\omega, k}^{i}}(t) +1\,)\,\log (\,n_{\mathcal{O}_{\omega, k}^{i}}(t) +1\,)\,-\,n_{\mathcal{O}_{\omega, k}^{i}}(t)\,\log (\,n_{\mathcal{O}_{\omega, k}^{i}}(t))\;.
\end{equation}

As for the toy model, the previous analysis shows the existence of three time scales. The first is the well-known local relaxation time scale, associated to quasinormal ringing. It is given by the inverse of the imaginary part of the quasinormal frequency, typically proportional to the inverse of the temperature. The second is the scrambling time scale, which again turns out to be independent of the system size $N$. It is controlled by the lowest quasinormal frequency. The third is the time by which deviations from equilibrium are uniform throughout the system. This time scale turns out to be:
\begin{equation}
t_{\textrm{random}}=\,\frac{1}{\Omega_{\mathcal{O}_{\omega, k}^{i}}^{^{\textrm{min}}}}\,\log N\;,
\end{equation}
where $\Omega_{\mathcal{O}_{\omega, k}^{i}}^{^{\textrm{min}}}$ is the imaginary part of the lowest quasinormal frequency. As we expand below, this time scale coincides with the time it takes a freely falling observer to cross the near horizon region until it hits the so-called brick wall \cite{brick} or stretched horizon \cite{complementarity}, as pointed out in \cite{us1,us3}.

In the last section, we comment on the information paradox \cite{paradox} in the light of our results. We show that the thermodynamic limit of democratic systems is essentially different from the thermodynamic limit of common local systems, being effectively described by non-unitary dynamics. The large-N limit acts as a coarse-graining by neglecting information originating in subleading correlation functions. It generates entropy without the need of tracing out part of the system. This observation naturally reconciles the old solution described in \cite{paradox} with unitary microscopic descriptions of black holes in which the interaction structure is democratic.

\section{Non-equilibrium unitary processes with random free fermions}\label{secII}

In this section, we study non-equilibrium dynamical processes using the family of Hamiltonians~(\ref{H}). Within the context of many-body quantum mechanics, there are many out of equilibrium initial conditions one can consider. Relaxation times, global or local (to be explicitly defined below), depend on the choice of the initial state. In this article, we consider two important classes of initial states.

The first initial state we consider intends to emulate the typical black hole thought experiment which consists of `throwing in' a small subsystem $\textbf{A}$ to a black hole. Subsystem $\textbf{A}$ qualifies as an `unentangled perturbation' of the black hole state. The combined system is left to relax through unitary evolution. Any information contained in $\textbf{A}$ is shared with the large number of internal degrees of freedom of the black hole and radiation \cite{hayden,susskind}, and becomes only accessible through fine-grained measurements of the global quantum state. Notice that the previous `natural' and `physical' assertion, concerning at the same time both information mixing and information conservation, has proven to be very difficult to observe in a specific model. There exist an obvious tension between the need of complicated dynamics to chaotically mix the initial information and the ability of actually solving the proposed model. As we show below, the family of Hamiltonians~(\ref{H}) overcome this tension.

The second initial state corresponds to a `collapse scenario', in which there is no black hole whatsoever to start with. In this context, an extreme situation arises by choosing as an initial state $N/2$ unentangled particles, a factorized state belonging to the highest entropic sector of the theory. In this case, the evolution of entanglement entropy is significant for any subsystem size.

\subsection{"Throwing in" an unentangled particle}

For the first out-equilibrium scenario, we will consider the initially unentangled subsystem to be the first oscillator. The initial quantum state is:
\begin{equation}\label{initial}
\rho_{\textrm{in}}=\rho_{1}\otimes\rho_{\beta}\;,
\end{equation}
defined by
\begin{equation}
\textrm{Tr}(\,\rho_{\textrm{in}} \,c^{\dagger}_{k}c_{l})=(n-f)\,\delta_{k1}\,\delta_{k1}+f\,\delta_{kl}\;.
\end{equation}
In the previous expressions, $\rho_{1}$ corresponds to the reduced density matrix of the first degree of freedom. It is itself a thermal density matrix with occupation number $n$. The rest of the degrees of freedom form the `black hole' and are set in the thermal ensemble $\rho_{\beta}=e^{-\beta H}/Z$, associated to the Hamiltonian~(\ref{H}). This thermal ensemble is fully characterized by its mean occupation number $f$. The computation of the average occupation number $f$ at temperature $\beta$ for the Hamiltonian~(\ref{H}) can be found in  \cite{useth}. Notice that gaussianity ensures that the two-point correlation function characterizes the full evolution of the state~(\ref{initial}), including entanglement entropies \cite{peschel}. Moreover, due to Wick's decomposition principle, stabilization of two-point correlation functions to the stationary value, whatever it is, implies stabilization of all higher point correlation functions as well. All time scales associated to the thermalization process are then encoded in the evolution of the time evolved two-point correlation matrix:
\begin{equation}
\mathcal{C}_{kl}(t)=\textrm{Tr}(\,\rho (t)\, c^{\dagger}_{k}c_{l})=\textrm{Tr}(\,\rho_{\textrm{in}}\, c^{\dagger}_{k}(t)c_{l}(t))\;,
\end{equation}
where $\rho (t) =U(t)\,\rho_{\textrm{in}}\,U^{\dagger}(t)$ and $c_{k}(t)=U^{\dagger}(t)\,c_{k}\,U(t)$. The unitary evolution operator is defined as usual by $U(t)=e^{-i\, H\, t}$.

To compute $\mathcal{C}_{kl}(t)$, notice that the `free' nature of the Hamiltonian~(\ref{H}) allows an exact solution via diagonalization of the matrix $V$. If $\psi^{a}$, for $a=1,\cdots , N$, are the eigenvectors of $V$ with eigenvalues $\epsilon_{a}$:
\begin{equation}
\sum\limits_{j=1}^{N}V_{ij}\,\psi^{a}_{j} = \epsilon_{a}\,\psi^{a}_{i}\;,
\end{equation}
and we define new creation and annihilation operators $d_{a}^{\dagger}$ and $d_{a}$ by:
\begin{equation}
d_{a}=\sum\limits_{i=1}^{N}\psi^{a}_{i}\,c_{i}\;,
\end{equation}
then the Hamiltonian can be equivalently written as:
\begin{equation}
H = \sum\limits_{a=1}^{N}(\alpha +\epsilon_{a})\, d_{a}^{\dagger}\,d_{a}= \sum\limits_{a=1}^{N}E_{a}\, d_{a}^{\dagger}\,d_{a}\;.
\end{equation}
In the basis $d_{a}$, the Hamiltonian is a set of decoupled fermionic oscillators. Their time evolution is:
\begin{equation}
d_{a}(t)=U^{\dagger}(t)\,d_{a}\,U(t)=e^{-i\,(\alpha+\epsilon_{b})\,t}d_{a}\;.
\end{equation}
Since $V$ is a random matrix, the creation and annhilitation operators create `random particles'. Their properties can be unravelled by using the theory of random matrices, which deals with the statistical properties of eigenvectors and eigenvalues of matrices such as $\eta V$. For an extensive treatment of random matrices see \cite{tao} and \cite{haake}. In relation to the eigenvalues, we will only need the widely known Wigner's semicircle law, accounting for the probability distribution of having an eigenvalue equal to $\lambda$. It reads:
\begin{equation}\label{wigner}
P(\lambda)=\frac{2}{\pi R^{2}}\,\sqrt{R^{2}-\lambda^{2}}\;,
\end{equation}
where $R^{2}\equiv 4N\eta^{2}$, and where we remind that it concerns the eigenvalues of $\eta\, V$, a matrix of size $N$ with deviation equal to $\eta$, see the Hamiltonian~(\ref{H}).

On the other hand, the statistical properties of eigenvectors are less widely known. The main assertion is that the orthogonal matrix of eigenvectors $(\psi^{1},\cdots ,\psi^{N} )$ is distributed according to the Haar measure on the orthogonal group $O(N)$. This means that the eigenvectors have independent and random gaussian entries, up to normalization:
\begin{equation}\label{statvec}
[\psi^{a}_{i}]=0 \,\,\,\,\,\,\,\, [\psi^{a}_{i}\,\psi^{b}_{j}]=\frac{1}{N}\,\delta_{ab}\,\delta_{ij}\;,
\end{equation}
wherein what follows we will use $[p]$ to denote the average of the random variable $p$ over the random matrix ensemble.

With this previous statistical information about eigenvalues and eigenvectors, we can compute the evolution of the correlation matrix $\mathcal{C}_{kl}(t)$ for a `typical' Hamiltonian belonging to the family~(\ref{H}). Notice that:
\begin{eqnarray}
&\mathcal{C}_{kl}(t)=\textrm{Tr}(\rho_{\textrm{in}} c^{\dagger}_{k}(t)c_{l}(t))=\sum\limits_{a,b=1}^{N}\psi^{a}_{i}\psi^{b}_{j}\,\textrm{Tr}(\rho_{\textrm{in}}d^{\dagger}_{a}(t)d_{b}(t))=&\nonumber \\ &\sum\limits_{a,b=1}^{N}\psi^{a}_{i}\psi^{b}_{j}\,e^{it(\epsilon_{a}-\epsilon_{b})}\,\textrm{Tr}(\rho_{\textrm{in}}d^{\dagger}_{a}d_{b})=\sum\limits_{k,l,a,b=1}^{N}\psi^{a}_{i}\psi^{b}_{j}\psi^{a}_{k}\psi^{b}_{l}\,e^{it(\epsilon_{a}-\epsilon_{b})}\,\textrm{Tr}(\rho_{\textrm{in}}c^{\dagger}_{a}c_{b})=&\nonumber \\ &\sum\limits_{k,l,a,b=1}^{N}\psi^{a}_{i}\psi^{b}_{j}\psi^{a}_{k}\psi^{b}_{l}\,e^{it(\epsilon_{a}-\epsilon_{b})}((n-f)\,\delta_{k1}\delta_{k1}+f\delta_{kl})&\;.
\end{eqnarray}
Defining
\begin{equation}
S_{ij}(t)\equiv\sum\limits_{a,b=1}^{N}\psi^{a}_{i}\,\psi^{b}_{j}\,\psi^{a}_{1}\,\psi^{b}_{1}\,e^{it(\epsilon_{a}-\epsilon_{b})}\;,
\end{equation}
the previous equation can be simply written as:
\begin{equation}\label{cor}
\mathcal{C}_{ij}(t)=\delta_{ij}\,f+(n-f)\,S_{ij}(t)\;.
\end{equation}
The intuition coming from the previous equation is the following. At the initial time $S_{ij}(t_{\textrm{in}})=\delta_{i1}\delta_{j1}$, we recover the unentangled initial state~(\ref{initial}), as we should. As time evolves $S_{ij}(t)\rightarrow 0$, and the final state is the global thermal distribution $\rho_{\beta}=e^{-\beta H}/Z$, characterized by its mean occupation number $f$.

The next step is to compute the statistical properties of $\mathcal{C}_{ij}(t)$. In terms of the statistical properties of $S_{ij}$, they are given by:
\begin{equation}
[\mathcal{C}_{ij}(t)]=\delta_{ij}f+(n-f)[S_{ij}(t)]\;,
\end{equation}
and
\begin{equation}
[\mathcal{C}_{ij}(t)^{2}]-[\mathcal{C}_{ij}(t)]^{2}=(n-f)^{2}([S_{ij}^{2}(t)]-[S_{ij}(t)]^{2})\;.
\end{equation}
To compute  the mean value $[S_{ij}(t)]$ and the deviations from the mean $(\sigma^{S}_{ij}(t))^{2}=[S_{ij}^{2}(t)]-[S_{ij}(t)]^{2}$ we use~(\ref{wigner}) and~(\ref{statvec}). We want to remark that if the deviations from the mean are small (as they will), we are indeed computing the typical properties of single Hamiltonian realizations taken from the family~(\ref{H}). This is due to self averaging, a known feature in random matrix theory.

The computation divides into different computations, as shown by the structure of the following matrix:
\begin{equation}
\mathcal{C}_{ij}(t)=
\begin{pmatrix}
\mathcal{C}_{11}(t)& \begin{pmatrix}\,\,\,
\cdots\,\,\,&\,\,\,\mathcal{C}_{1j}(t)\,\,\,&\,\,\,\cdots\,\,\,\\
\end{pmatrix}\\[-3mm] \\
\begin{pmatrix}
\vdots\\ \mathcal{C}_{1j}^{*}(t)\\ \vdots\\
\end{pmatrix}&\begin{pmatrix}\ddots& & \mathcal{C}_{ij}(t)\\ &\mathcal{C}_{ii}(t)& \\ \mathcal{C}_{ij}^{*}(t)& &\ddots\\ \end{pmatrix}\\
\end{pmatrix}\;,
\end{equation}
where we mean that $\mathcal{C}_{1j}(t)=\mathcal{C}_{1k}(t)$, $\mathcal{C}_{jj}(t)=\mathcal{C}_{kk}(t)$ and $\mathcal{C}_{ij}(t)=\mathcal{C}_{kl}(t)$ for $j,k,i,l\neq 1$. Let's start by $S_{11}(t)$. The mean is given by:
\begin{equation}
[S_{11}(t)]=[\sum\limits_{a,b=1}^{N}\psi^{a}_{1}\psi^{b}_{1}\psi^{a}_{1}\psi^{b}_{1}\,e^{it(\epsilon_{a}-\epsilon_{b})}]=\sum\limits_{a,b=1}^{N}[\psi^{a}_{1}\psi^{b}_{1}\psi^{a}_{1}\psi^{b}_{1}]\,[e^{it(\epsilon_{a}-\epsilon_{b})}]\;,
\end{equation}
where in the second line we have used the fact that eigenvenctors and eigenvalues are not correlated in the large $N$ limit. The leading term comes from the following contraction $[\psi^{a}_{1}\psi^{a}_{1}]\,[\psi^{b}_{1}\psi^{b}_{1}]$, since in this case we do not kill any sum. Noticing that the $a=b$ part of the sum is time independent:
\begin{equation}
[S_{11}(t)]=\frac{3}{N}+\sum\limits_{a\neq b}\frac{1}{N^{2}}\,([\,\cos(t\,(\epsilon_{a}-\epsilon_{b}))\,]+i\,[\,\sin(t\,(\epsilon_{a}-\epsilon_{b}))\,]\,)\;.
\end{equation}
The averages over the spectrum are given by:
\begin{equation}
[\,\cos(t(\epsilon_{a}-\epsilon_{b}))\,]=\int\limits_{-R}^{R} d\epsilon_{a}d\epsilon_{b}\,\frac{4}{\pi R^{4}}\,\sqrt{R^{2}-\epsilon_{a}}\,\sqrt{R^{2}-\epsilon_{b}}\cos(t(\epsilon_{a}-\epsilon_{b}))=\frac{4\,(J_{1}(Rt))^{2}}{(Rt)^{2}}\;,
\end{equation}
where $J_{1}(x)$ is the Bessel function of the first kind. The final result reads:
\begin{equation}
[S_{11}(t)]=\frac{3}{N}+\frac{N(N-1)}{N^{2}}\frac{4\,(J_{1}(Rt))^{2}}{(Rt)^{2}}\simeq \frac{3}{N}+\frac{4\,(J_{1}(Rt))^{2}}{(Rt)^{2}}\;.
\end{equation}
We remind that the Bessel function $J_{1}(Rt)$ is an oscillatory decaying function. For large argument $R\,t\gg 1$, the previous expression behaves as:
\begin{equation}
[S_{11}(t\gg 1/R)]\simeq \frac{3}{N}+\frac{8}{\pi (Rt)^{3}}\cos^{2} (Rt-\frac{3\pi}{4})\;.
\end{equation}
By an analogous procedure, we can compute all the other mean values and deviations. The computations are somewhat tedious, especially for $[S_{11}(t)^{2}]$, so we will just quote the results in matrix form. The mean is given by:
%\begin{eqnarray}
%[S_{11}(t)]&=&\frac{3}{N}+\frac{4(J_{1}(Rt))^{2}}{t^{2}}\nonumber\\
%(\sigma^{S}_{11}(t))^{2}&=&\frac{8}{N^{2}}+\frac{16}{N}\frac{(J_{1}(Rt))^{2}}{t^{2}}(1+\frac{J_{1}(2Rt)}{t})\nonumber\\
%[S_{11}(t)]&=&
%\end{eqnarray}

\begin{equation}\label{S}
[S_{ij}(t)]=
\begin{pmatrix}
\frac{3}{N}+\frac{4\,(J_{1}(Rt))^{2}}{(Rt)^{2}}& \begin{pmatrix}
\,\,\cdots&0&\cdots\,\,\\
\end{pmatrix}\\[-3mm] \\
\begin{pmatrix}
\vdots\\ 0\\ \vdots\\
\end{pmatrix}&\begin{pmatrix}\ddots& & 0\\ &0& \\ 0& &\ddots\\ \end{pmatrix}\\
\end{pmatrix}\;,
\end{equation}
while the deviation reads:
\begin{equation}\label{dS}
[(\sigma^{S}_{ij}(t))^{2}]=
\begin{pmatrix}
\frac{8}{N^{2}}+\frac{16}{N}\frac{(J_{1}(Rt))^{2}}{(Rt)^{2}}(1+\frac{J_{1}(2Rt)}{Rt})& \begin{pmatrix}
\,\,\cdots&\frac{15}{N^{3}}+\frac{4(J_{1}(Rt))^{2}J_{1}(2Rt)}{N\,(Rt)^{3}}&\cdots\,\,\\
\end{pmatrix}\\[-3mm] \\
\begin{pmatrix}
\vdots\\ \frac{15}{N^{3}}+\frac{4(J_{1}(Rt))^{2}J_{1}(2Rt)}{N\,(Rt)^{3}}\\ \vdots\\
\end{pmatrix}&\begin{pmatrix}\ddots& & \frac{3}{N^{3}}+\frac{4(J_{1}(2Rt))^{2}}{N^{2}\,(Rt)^{2}}\\ &\frac{2}{N^{2}}+\frac{4(J_{1}(2Rt))^{2}}{N^{2}\,(Rt)^{2}}& \\ \frac{3}{N^{3}}+\frac{4(J_{1}(2Rt))^{2}}{N^{2}\,(Rt)^{2}}& &\ddots\\ \end{pmatrix}\\
\end{pmatrix}\;.
\end{equation}
Joining all results together, we can write the evolution of the correlation matrix $\mathcal{C}_{ij}(t)$ schematically as:
\begin{equation}\label{cort}
\mathcal{C}_{ij}(t)=f\delta_{ij}+(n-f)\begin{pmatrix}
\frac{4\,(J_{1}(Rt))^{2}}{(Rt)^{2}}\pm\mathcal{O}(1/N)& \begin{pmatrix}
\,\,\,\,\,\,\,\,\,\cdots\,\,\,\,\,\,\,&\pm \mathcal{O}(1/N^{\frac{3}{2}})&\,\,\,\,\,\,\,\cdots\,\,\,\,\,\,\,\,\,\\
\end{pmatrix}\\[-3mm] \\
\begin{pmatrix}
\vdots\\ \pm \mathcal{O}(1/N^{\frac{3}{2}})\\ \vdots\\
\end{pmatrix}&\begin{pmatrix}\ddots& & \pm \mathcal{O}(1/N^{\frac{3}{2}})\\ &\pm\mathcal{O}(1/N)& \\ \pm \mathcal{O}(1/N^{\frac{3}{2}})& &\ddots\\ \end{pmatrix}\\
\end{pmatrix}\;.
\end{equation}
From these formulas we conclude that the initial out of equilibrium state~(\ref{initial}) thermalizes and evolves towards a state with homogeneous average occupation given by $f$. The final state is a `global black hole'.

Besides, since the system is gaussian, the entanglement entropy of any subsystem $A$ can be computed using the framework developed in \cite{peschel}. This framework writes the entanglement entropy in terms of the eigenvalues of the correlation matrix of subsystem $A$. If $\mathcal{C}_{ij}$, for $i,j \in A$, is such a correlation matrix and $\lambda_{m}$ are its eigenvalues, the entanglement entropy can be written as:
\begin{equation}
S_{A}=-\textrm{Tr}(C\log C + (1-C)\log (1-C))=-\sum\limits_{m}(\lambda_{m}\log \lambda_{m} + (1-\lambda_{m})\log (1-\lambda_{m}))\;.
\end{equation}
The key point to observe now is that off-diagonal terms in the correlation matrix~(\ref{cort}) vanish in the thermodynamic limit. Consider a subsystem $A$ composed of $M$ oscillators. We can express the correlator matrix as:
\begin{equation}\label{cex}
C_{ij}=n_{i}(t)\delta_{ij}+(\Delta C)_{ij}\;,
\end{equation}
where $n_{ij}(t)$ is the occupation density of the oscillator $i$, and $\Delta C$ is a random matrix taken from the GOE ensemble with zero mean and deviation $\sigma_{C}\sim \mathcal{O}(1/N^{3/2})$. To compute the entropy we need to compute:
\begin{equation}\label{averageent}
[S_{A}]=-[\textrm{Tr}(C\log C + (1-C)\log (1-C))]\;.
\end{equation}
This can be approximately computed by using the techinques of Appendix A in \cite{usrandom}. The trace is a sum over the eigenvalues of $C$. These can be expressed as $C_{i}=n_{i}+(\Delta C)_{i}$, where $(\Delta C)_{i}$ are the eigenvalues of $\Delta C$. Using the semicircle law~(\ref{wigner}), we obtain $[(\Delta C)_{i}]=0$ and $[(\Delta C)_{i}^{2}]\propto M/N^{3/2}$. Taylor expanding~(\ref{averageent}) we obtain:
\begin{equation}\label{averageent2}
[S_{A}]=-\sum\limits_{i=1}^{M}(n_{i}(t)\log n_{i}(t) + (1-n_{i}(t))\log (1-n_{i}(t)))-\mathcal{O}(\frac{M^{2}}{N^{3/2}})\;.
\end{equation}
We conclude that the Von Neumann entropy in the thermodynamic limit is extensive at all times, even for big subsystems. For a susbystem $B$ not including the first degree of freedom it is just the thermal entropy of $B$, independent of time. For a subsystem $B\equiv 1\cup A$ including the first degree of freedom, it is just the thermal entropy of $A$ plus $S_{1}(t)$. Therefore, in this scenario, the most interesting entanglement entropy is the one associated to the first degree of freedom. It is just given by:
\begin{equation}\label{entt}
S_{1}(t)=-\textrm{Tr}(\,\rho_{1}(t)\,\log \rho_{1}(t)\,)=-\mathcal{C}_{11}(t)\log\mathcal{C}_{11}(t)-(1-\mathcal{C}_{11}(t))\log(1-\mathcal{C}_{11}(t))\;.
\end{equation}
Using~(\ref{cort}) we observe that $S_{1}(t)$ evolves towards the thermal entropy of the Fermi distribution, saturating at a distance of $\mathcal{O}(1/N)$ from it. Notice that stationarity, the time in which the initial factorization of the density matrix~(\ref{initial}) is broken, is reached in a time scale $t_{\textrm{r}}=1/R$ independent of the system size. It is clear from this result that mean field approximations of democratic systems in time dependent scenarios have a very limited range of applicability. We will expand more on this issue and on its applications to black hole physics below.

The previous equations~(\ref{S}),~(\ref{dS}),~(\ref{cort}), and~(\ref{entt}) constitute the main results of this section. In section~(\ref{secIII}) we will use them to make several remarks about information spreading in democratic systems, and comment on their applications to black hole physics.

\subsection{Entanglement dynamics from factorized initial states}

The second class of initial states we consider is the class of completely factorized states. More concretely, we can have an initial state in which the first $\bar{N}$ oscillators are excited, while the $N-\bar{N}$ left over are not:
\begin{equation}\label{factorized}
\vert\Psi_{\textrm{in}}\rangle=c^{\dagger}_{1}\,c^{\dagger}_{2}\,\cdots\, c^{\dagger}_{\bar{N}}\,\vert 0\rangle\;.
\end{equation}
This situation emulates a `collapse' scenario, in which $\bar{N}$ initially unentangled particles interact with each other, forming a black hole at late times. To analyze the time evolution we need to write the state in terms of the decoupled oscillator basis $d_{a}^{\dagger}$:
\begin{equation}\label{evolvedst}
\vert\Psi (t)\rangle= \sum\limits_{j,k,\cdots ,l}\psi^{j}_{1}\psi^{k}_{2}\cdots \psi^{l}_{\bar{N}}\,e^{i\,t\,(E_{j}+E_{k}+\cdots +E_{l})}d^{\dagger}_{j}\,d^{\dagger}_{k}\cdots d^{\dagger}_{l}\,\vert 0\rangle\;.
\end{equation}
The time evolved correlator matrix is then:
\begin{eqnarray}
\mathcal{C}_{ij}(t)&=&\langle\Psi (t)\vert \,c_{i}^{\dagger}c_{j}\,\vert\Psi (t)\rangle \nonumber \\ &=&\sum\limits_{j,l;k,m;n,p}\psi^{j}_{1}\cdots \psi^{l}_{\bar{N}}\psi^{k}_{1}\cdots \psi^{m}_{\bar{N}}\psi^{n}_{i}\psi^{p}_{j}\,e^{-i\,t\,(E_{j}+\cdots +E_{l}-E_{k}-\cdots -E_{m})}\times\nonumber\\ &\times&\langle 0\vert \,d_{j}\,\cdots d_{l}\,d^{\dagger}_{n}\,d_{p}\,d^{\dagger}_{k}\,\cdots \,d^{\dagger}_{m}\,\vert 0\rangle\;.
\end{eqnarray}
The computation is severely more complicated. There is an exponentially growing number of terms, with respect to the number of initial particles. This is due to the vacuum ($2\bar{N}+2$)-point correlation function, after applying Wick's decomposition principle.

Although we were not able to compute the full statistical properties of the previous quantity, it is possible to observe that it is diagonal on average:
\begin{equation}\label{caver}
i\neq j\longrightarrow [\,\mathcal{C}_{ij}(t)\,]=0\;.
\end{equation}
On average, the only surviving entries are the diagonal ones, corresponding to the expectation values of the number operators. This could have been expected from generic considerations, since the process must be similar to the previous case, in regards to information spreading. The intuition is the following. Given the randomness of $V$ in the Hamiltonian~(\ref{H}), from the point of view of a single degree of freedom, the rest of the system behaves as a thermal bath at all times. Relation~(\ref{caver}) is an example of large-$N$ factorization in fully time-dependent scenarios.

Since there is an effective permutation symmetry between the degrees of freedom in the Hamiltonian~(\ref{H}), the decay of the number operator of the initially excited particles is the same for all of them $[\langle c_{\uparrow}^{\dagger}c_{\uparrow}\rangle] (t)= n_{\uparrow} (t)$. The same can be said about the decays of the number operator associated to the 
oscillators that were initially non-excited $[\langle c_{\downarrow}^{\dagger}c_{\downarrow}\rangle (t)]= n_{\downarrow} (t)$. The average value of the correlator matrix has then the following diagonal form:
\begin{equation}\label{diagonalC}
[\,\mathcal{C}_{ij}(t)\,]=\textrm{Diag} (\,n_{\uparrow} (t),\cdots,n_{\uparrow} (t),n_{\downarrow} (t),\cdots,n_{\downarrow} (t)\,)\;,
\end{equation}
where there are $\bar{N}$ entries with $n_{\uparrow} (t)$ and $N-\bar{N}$ entries with $n_{\downarrow} (t)$.

We will assume in what follows that the off-diagonal deviations $\sigma_{C}^{2}=[C_{ij}^{2}]-[C_{ij}]^{2}$ have the same structure as in the previous case, and are of $\mathcal{O}(1/N^{ p})$, for some positive $p$ \footnote{Indeed, from the results of the previous section we expect $p= 3$. If we take the static results of \cite{useth} we expect at least $p\geq 1$. For large-N matrix models typically we have $p\geq 1$. The ultimate reason for the suppression is the well-known fact that the connected part of a correlator in a large-N theory kills at least one sum more}. This is a natural assumption given the results of the previous section and the intuition coming from large-N factorization in matrix models. At any rate, we stress that the following statements rely on such assumption.

If such assumption is correct, we can repeat the computation of the previous section, formulas~(\ref{cex}) and~(\ref{averageent2}). Considering a subsystem $A$ composed of $M$ oscillators we obtain:
\begin{equation}
[S_{A}]=-\sum\limits_{i=1}^{M}[n_{i}(t)\log n_{i}(t) + (1-n_{i}(t))\log (1-n_{i}(t))]-\mathcal{O}(\frac{M^{2}}{N^{p/2}})\;.
\end{equation}
From the previous relations we conclude that the deviation from entanglement extensivity is subleading for \emph{any} $M\lesssim N/2$ if $p\geq 2$, which is a natural expectation as commented before. Notice that for $M=a N$, the leading term is proportional to $a$ while the subleading term is proportional to $a^{2}$.

Therefore, up to subleading corrections, the average entanglement entropy of a subset of $M=M_{\uparrow}+M_{\downarrow}$ oscillators satisfy the following scaling law:
\begin{equation}\label{scaling}
S_{M}=M_{\uparrow}\,S_{M_{\uparrow} }(t)+M_{\downarrow}\,S_{M_{\downarrow} }(t)\;,
\end{equation}
where
\begin{equation}\label{scaling1}
S_{n_{\uparrow} }(t)=-n_{\uparrow} (t)\,\log n_{\uparrow} (t)-(1-n_{\uparrow} (t)\,)\log(\,1-n_{\uparrow} (t)\,)\;,
\end{equation}
is the entanglement entropy associated to one oscillator that was initially excited, and where a similar relation holds for $S_{n_{\downarrow}} (t)$. The previous scaling relation was found by generic arguments concerning information spreading in democratic systems in \cite{uswalks}. It seems to be a distinctive feature of democratic systems, when compared to local ones. Relation~(\ref{scaling}) shows that the global properties of information spreading, such as the growth of entanglement entropies of large subsystems are fully controlled by the growth of entanglement entropies of each oscillator separately. But from~(\ref{scaling}) we gain a further insight. Since the single particle entanglement entropies are themselves controlled by the decay of the number operators $n (t)$, the global properties of information spreading are encoded in the local relaxation properties as well.

Given the previous results, it is of obvious interest to find $n(t)$ as a function of the $\bar{N}$, in particular its characteristic time scale. This can be computed without too much trouble for the case of one excitation $\bar{N}=1$, in which:
\begin{equation}\label{N1}
[\,\langle \,c_{\uparrow}^{\dagger}c_{\uparrow}\,\rangle\,]\,(t)=\frac{4\, (J_{1}(Rt))^{2}}{(Rt)^{2}}\;,
\end{equation}
decaying to zero at large times. This is expected since the final state has negligible energy. The entanglement entropy associated to the first oscillator grows at intial times, attains its maximum possible value for $[\,\langle \,c_{\uparrow}^{\dagger}c_{\uparrow}\,\rangle\,]\,=1/2$ and then decays to zero for times $t\gg 1/R$. Notice again that this simple calculation shows that entanglement is created in times of $\mathcal{O}(1/R)$. Mean field methods fail to explain the dynamics at these early time scales.

Also, notice that the simple computation of $\bar{N}=1$ could have been predicted with the solution of the previous section. Namely, if we take~(\ref{cort}) and set $f=0$ and $n=1$ we obtain~(\ref{N1}) as the leading term in the thermodynamic expansion. This suggests that each degree of freedom sees the rest of the system as a thermal bath already by times $t\ll 1/R$ much shorter than the relaxation time scale. Looking at the evolved state~(\ref{evolvedst}), we might expect that the state becomes similar to the random state in the $\bar{N}$-particle sector\footnote{For a characterization of random states in sectors with a fixed number of particles see \cite{useth}.} when the phases randomize. This gives a time scale  of the order of the inverse of the energy sum in~(\ref{evolvedst}). In turn, due to the central limit theorem, the energy sum is a gaussian random variable with squared deviation given by $\sigma^{2}_{\sum E}=\bar{N}R^{2}$. The time scale for dephasing is $Rt\sim\mathcal{O}(1/\sqrt{\bar{N}})$, dying in the thermodynamic limit. It seems natural to expect that at very early times $t\sim R/\sqrt{\bar{N}}$, each degree of freedom sees the rest of the system as a thermal bath, and, therefore, decays according to~(\ref{cort}) \footnote{The previous assertion needs to be rigorously proven. Indeed, the previous argument might be supporting the claim that the global time scale for relaxation is $Rt\sim\mathcal{O}(1/\sqrt{\bar{N}})$ in these collapse scenarios. We leave this issue for future work.}

Finally, notice that the results of this section are not expected to depend on the initial factorized state we consider. For a generic factorized state of $N$ fermions:
\begin{equation}\label{factgen}
\vert\Psi_{\textrm{in}}\rangle=\vert\psi_{1}\rangle\otimes\vert\psi_{2}\rangle\otimes\cdots\otimes\vert\psi_{N}\rangle;,
\end{equation}
one can always make a set of local unitary transformations and bring the state towards the form~(\ref{factorized}). Of course, this is not a symmetry of the Hamiltonian~(\ref{H}). But given that the Hamiltonian is random, we know that the eigenstates are `non-locally' spread through the whole system, see relation~(\ref{statvec}). Given that we have preformed a set of local unitary transformations, the Hamiltonian in the new basis still randomly connects degrees of freedoms at different sites. This again implies large-$N$ factorization~(\ref{caver}) for the initial state~(\ref{factgen}), and the evolution of entanglement entropy of a set $A$ of degrees of freedom is expected to be generically given  by:
\begin{equation}\label{fermions}
S_{A}^{\textrm{fermions}}=-\sum\limits_{i\in A}(\,n_{i}(t)\log n_{i}(t) + (1-n_{i}(t)\,)\log (\,1-n_{i}(t))\,)\;.
\end{equation}
Finally, as commented in the introduction, we could have well considered bosons instead of fermions.  The only thing that would change is the relation between the entropy and the number operators. In the case of bosons it would read:
\begin{equation}\label{bosons}
S_{A}^{\textrm{bosons}}=\sum\limits_{i\in A}(\,(n_{i}(t)+1)\log (\,n_{i}(t)+1)-n_{i}(t)\,\log n_{i} (t)\,)\;.
\end{equation}
Relations~(\ref{fermions}) and~(\ref{bosons}) have been seen to appear in \cite{useth,usrandom} when considering eigenstates of fully random Hamiltonians or eigenstates of random free fermions. They are rooted in the microscopic non-locality of the quantum state, allowing the computation of entanglement entropies for small enough subsystems solely by means of the diagonal entries of the reduced density matrices. These diagonal entries are the expectation values of the number operators. What relations~(\ref{cort}),~(\ref{caver}),~(\ref{fermions}) and~(\ref{bosons}) show is that these relations are not only valid at stationary states. Due to the democratic structure of interactions, these relations are valid at all times in the thermodynamic limit.

\section{Black hole physics, random particles and matrix models}\label{secIII}

In this section we apply the results of the previous section, concerning the dynamical behavior of random particles, to the physics of black holes and large-$\mathcal{N}$ Matrix models. These large-$\mathcal{N}$ Matrix models hold key clues towards understanding the emergence of black hole geometric backgrounds \cite{matrix,adscft,papadodimas}, and we would like to understand in more detail several aspects of their color dynamics.

One crucial aspect to analyze in this regard is how perturbations (and the information associated with them) spread through the system as time evolves \cite{susskind}. A related aspect of primary interest in the community is to study these thermalization processes through entanglement entropy \cite{esperanza,vijay} since entanglement entropy has a direct geometrical meaning \cite{takayanagi}. Finally, it is interesting to see what are the peculiarities of democratic models in regards to the information paradox \cite{paradox}.

In this section we analyze aspects of those three questions. We first do it for the random particle toy model, the main objective being to gain the necessary intuition in a simplified solvable scenario. Besides, as we will see, the results are interesting in their own right. Later on, we show how the intuition naturally extends to large-$\mathcal{N}$ matrix models, which furnish exact conjectured descriptions of quantum black holes.

\subsection{Information transport and scrambling with random particles}

In the context of black hole physics, information transport in thermal processes was coined `information scrambling' in \cite{hayden,susskind}. A lower bound was conjectured for such information scrambling in physical systems:
\begin{equation}
t_{\textrm{scrambling}}\sim \beta \,\log N\;,
\end{equation}
where $\beta$ is the temperature of the system or other significant time scale associated with the Hamiltonian or initial state considered, and $N$ the number of black hole degrees of freedom.

Let's begin with defining in various different ways \emph{information scrambling}. In \cite{susskind} it was broadly defined as the time scale by which a given perturbation spreads or contaminates the whole system. In local theories the intuition leads naturally to consider diffusion processes. It was then more rigorously (and abstractly) defined by the `Page's test' \cite{page}, as the time by which subsystems are maximally entangled with the environment, see \cite{susskind,lashkari,us3}. The logic behind this last definition is well known in the context of quantum thermalization, see the recent review \cite{eisertreview}. If we choose a generic subsystem $A$ smaller than half of the system, with reduced density matrix $\rho_{A}$, the non-equilibrium unitary process will drive this reduced state towards the thermal distribution at temperature $\beta$. This implies that the entanglement entropy associated to $A$ runs towards the thermal entropy as well. Asking for such an entanglement relaxation to all possible subsystems is a global notion of thermalization. The problem with this abstract notion of thermalization is that its connection to actual `information transport' is to some extent opaque. This gap was filled in \cite{uscod,uswalks}, in which information transport was carefully defined by the evolution of the structure of correlations between subsystems in the global state. In this setup, the scrambling time is naturally defined by the time scale in which correlations uniformize in size so that information is transported to the whole system democratically. This democratic structure of correlations is a characteristic feature of random pure states \cite{uscod,usrandom}.

We can analyze most types of definitions with the global solution~(\ref{cort})\footnote{Let us stress a technical issue at this point. The fact that we perturb only one degree of freedom~(\ref{initial}) does not mean we cannot discuss information scrambling. Information about the first degree of freedom, i.e its initial occupation number, is shared with the rest of the system as time evolves. This spreading can be defined more properly by using Mutual Information, as was layed out in \cite{uscod}.}. The first thing we observe is that correlations spread instantaneously. From~(\ref{cort}) it is clear that any property of global relaxation can be characterized by analyzing the number operator associated to the first degree of freedom:
\begin{equation}
\mathcal{C}_{11}(t)\simeq f+(n-f)\,(\,\frac{4\,(J_{1}(Rt))^{2}}{(Rt)^{2}}\pm\mathcal{O}(1/N)\,)\;.
\end{equation}
In $\mathcal{C}_{11}(t)$ appears the essential time scale characterizing information spreading, which is given by $t_{\textrm{relax}}=1/R$. Indeed, in the thermodynamic limit:
\begin{equation}\label{diagonalCt}
C_{ij}(\,t\gg 1/R\,)= f\,\delta_{ij}+(\,\mathcal{C}_{11}(t)-f\,)\,\delta_{i1}\,\delta_{j1}\;.
\end{equation}
For $t\gg 1/R$ the correlation matrix is simply the global thermal density matrix:
\begin{equation}\label{diagonalCt2}
C_{ij}(\,t\gg 1/R\,) \simeq f\,\delta_{ij}\;,
\end{equation}
up to computed subleading corrections. The information lost by the first degree of freedom is instantaneously spread through the whole system, since correlations are uniform through the system $C_{1j}(t)\sim C_{1k}(t)$. This is of course due to the microscopic non-locality.

The fact that correlations are at all times uniform over the system does not immediately imply scrambling. A clear and extreme example of this situation arises when all correlations are zero, as happens for the initial state considered before:
\begin{equation}\label{initial2}
\rho_{\textrm{in}}=\rho_{1}\otimes\rho_{\beta}\;,
\end{equation}
where there is a clear factorization between the first degree of freedom and the thermal ensemble.  Scrambling of the information associated with the first degree of freedom requires correlations to be uniform over the system \emph{plus} the first degree of freedom to be maximally entangled with the environment. Since any created correlations are homogeneously spread through the system, the time of entanglement production, or analogously the time in which the initial factorization~(\ref{initial2}) breaks down \emph{is} the scrambling time itself. 

The entanglement entropy was computed in the previous section to be:
\begin{equation}\label{sc11}
S_{1}(t)=-\mathcal{C}_{11}(t)\log\mathcal{C}_{11}(t)-(1-\mathcal{C}_{11}(t)\,)\,\log(1-\mathcal{C}_{11}(t)\,)\;,
\end{equation}
which takes the Von Newmann entropy at the initial time:
\begin{equation}
S(t_{\textrm{in}})=-(\,n\log n + (1-n\,)\,\log (1-n\,)\,)\;,
\end{equation}
to the Von Newmann entropy associated to the global thermal state:
\begin{equation}
S(t\gg t_{\textrm{local}})=-(\,f\log f + (1-f\,)\,\log (1-f\,)\,)\;.
\end{equation}
The characteristic time scale for stationarity of entanglement entropy is thus $t_{\textrm{local}}=1/R$. Also, as shown in the previous section, notice that any other entanglement entropy associated to other subsystems reach the entanglement plateaux at this time scale as well.

At this point, we want to remark that this result, if correct, furnish a counterexample of the mean field bound presented in \cite{lashkari}. In Ref. \cite{lashkari} it was claimed that there is a lower bound for the time scale of entanglement production in gaussian systems with fully non-local interactions. This lower bound was claimed to be proportional to $\log N$. More specifically, it is claimed that for models such as~(\ref{H}) the time evolution of the $N$ oscillator system can be approximated by:
\begin{equation}\label{mf}
U(t)=e^{-i H t}\simeq \prod\limits_{i=1}^{N}U^{\textrm{MF}}_{i}(t)\;,
\end{equation}
for some suitably defined `mean field local unitary evolution' $U^{\textrm{MF}}_{i}(t)$, acting in each oscillator separately. The claim is that this approximation is valid until times proportional to $\log N$. If we would apply such approximation to our initial state~(\ref{initial}), it would imply that until such large times entanglement could not be created. We believe that our results, formulas~(\ref{cort}),~(\ref{entt}) and~(\ref{N1}), which follow from relatively simple computations, show that production and stationarity of entanglement entropy are attained in a time scale independent of the system size. If correct, they furnish a counterexample to the arguments presented in \cite{lashkari}, since this early time entanglement cannot be explained with a mean field approximation, such as~(\ref{mf}). It is obvious that this issue deserves further consideration. Finally, in view of our results, notice that there is indeed some mean field approximation applying here. In the large-N limit, and before times of $\mathcal{O}(\beta \log N)$, the evolution of the global state  can be approximated by:
\begin{equation}\label{mmf}
\rho (t)\simeq \prod\limits_{i=1}^{N}\rho_{i}(t)\;,
\end{equation}
where $\rho (t)$ is not obtained from $\rho (0)$ by unitary evolution but by some super-evolution operator which change the on-site Von Neumann entropy of the reduced on-site state. This is another path to understand~(\ref{fermions}) and~(\ref{bosons}).

Coming back to scrambling, and to make the statement even clearer, we can also use the approach to information scrambling proposed in \cite{uscod,uswalks}.
Due to the effective permutation symmetry of the random free fermions Hamiltonian~(\ref{H}), for any pair of subsystems $A$ and $B$ with the same size, the Mutual Information between the first degree of freedom and those subsystems is equal:
\begin{equation}\label{spread}
[\,I(1,A)(t)\,]\equiv[\,S_{1}+S_{A}-S_{1A}\,]=[\,I(1,B)(t)\,]=[\,I(1,B)(t)\,]\equiv [\,S_{1}+S_{B}-S_{1B}\,]\;.
\end{equation}
For a subsystem $A$ of $M$ degrees of freedom, using~(\ref{averageent2}) we obtain:
\begin{equation}\label{mutualvanish}
I(1,A)(t)\sim \mathcal{O}(M^{2}/N^{3/2})\;,
\end{equation}
We now join all features together. The information lost by the first degree of freedom, quantified by $\Delta S_{1}(t)$, satisfies two properties. First, it is non-locally spread through the whole system, the Mutual Information being of the same size for any given pair of subsystems with the same size~(\ref{spread}). This is due to the democratic structure of interactions in the microscopic Hamiltonian~(\ref{H}). Second, this information can only be recovered by looking to big enough subsystems, since for small enough subsystems the Mutual Information vanishes in the thermodynamic limit~(\ref{mutualvanish}). It is of obvious interest to quantify how big enough the subsystem must be to recover the information about the first degree of freedom. However, we want to stress that subsystems with a size not scaling with $N$ in the thermodynamic limit will not achieve this goal. From this precise point of view, the information lost by the first degree of freedom is fully scrambled over the environment. Given~(\ref{sc11}), by $t_{\textrm{local}}=1/R$ the first degree of freedom has shared an $\mathcal{O}(1)$ amount of its information content with the environment. Therefore, after $t_{\textrm{local}}=1/R$ has elapsed, an $\mathcal{O}(1)$ amount of information has been shared non-locally with an $\mathcal{O}(N)$ amount of degrees of freedom.

We can repeat all the arguments for the factorized initial state~(\ref{factorized}). The only thing we need to assume to arrive at the same set of conclusions is again that the mean squared deviation of the correlation matrix is of $\mathcal{O}(1/N^{p})$, for $p\geq 1$. As explained in the previous section this is a natural assumption given the large-N matrix structure of the model and given the exact results obtained with the first initial state. If such assumption is correct, entanglement entropies evolve extensively, even for subsystems with extensive size $M\lesssim N/2$. This implies that saturation at the plateaux of the subsystem occurs when one degree of freedom saturates by itself, i.e the time in which the on-site entanglement is created. This is equivalent to the time by which the mean field approximation breaks down, which is just given by the local relaxation time scale.

Summarizing, from the behavior of entanglement entropies, Mutual Information and correlation functions, we conclude that the time scale associated to:
\begin{itemize}
\item Information lost by on-site degrees of freedom, quantified by their associated entanglement entropy,
\item Information spreading through the black hole degrees of freedom, characterized by correlations and Mutual Information being uniform through the system,
\item Relaxation of all entanglement entropies to thermal values at leading order,
\item Breakdown of the mean field approximation
\end{itemize}
is size independent and given by:
\begin{equation}\label{timescale}
t_{\textrm{local}}=t_{\textrm{scrambling}}=1/R\;.
\end{equation}
We conclude that the family of Hamiltonians~(\ref{H}) studied in this article furnish an specific example of the generic statement described in \cite{uswalks}:
\begin{itemize}
\item For democratic systems, information spreads instantaneously. The characteristic time scales appear already at the level of local relaxation of perturbations.
\end{itemize}
On the other hand, these results challenge the fast scrambling conjecture \cite{susskind}. The fast scrambling conjecture states that the time scale for a global spreading of information is bounded from below by $t_{\textrm{scrambling}}\gtrsim \beta\log N$, while here we find a time scale independent of the system size. We want to stress again that the lower bounds presented in Ref.\cite{lashkari} seem not to be valid generically. The breakdown of the mean field approximation, or analogously the breakdown of the factorization of the initial state~(\ref{initial2}), occurs on a time scale given by $t=1/R$, independent of the system size, see~(\ref{sc11}). In such a situation, given that the spreading of information is instantaneous, the model~(\ref{H}) seem to furnish a counterexample of the claimed fast scrambling lower bound.

\subsubsection{Scrambling vs randomization time scales}

In the previous section we have shown that an $\mathcal{O}(1)$ amount of information about the initial state is spread in a time of $\mathcal{O}(1/R)$ over the entire system. But has the thermalization process reach equilibrium in all possible senses?

Looking to the exact form of the solution~(\ref{S}),~(\ref{dS}),(\ref{cort}), we see that after the relaxation time scale $\mathcal{O}(1/R)$ has elapsed, there is still a long way towards `\emph{global equilibration of the deviations from thermality}'. To be precise, the set of number operators corresponding to the black hole internal degrees of freedom (the oscillators with $i\neq 1$), are shifted from the thermal expectation value by:
\begin{equation}
\mathcal{C}_{ii}-f\sim \mathcal{O}(1/N)\;,
\end{equation}
already at times of $\mathcal{O}(1/R)$. On the other hand, the deviation from thermality of the number operator corresponding to the first degree of freedom is:
\begin{equation}\label{c11}
\mathcal{C}_{11}-f =(n-f)\, \frac{4\,(J_{1}(Rt))^{2}}{(Rt)^{2}}\pm \mathcal{O}(1/N)\;.
\end{equation}
It is now natural to define a `randomization' time scale $t_{\textrm{random}}$, by asking the deviations from thermality to be uniform over the system. Notice that this is not equivalent to ask for correlations to be uniform over the system. From a more physical perspective, it is defined as the time scale at which the leading term in~(\ref{c11}), the term driving the decay, ceases to be valid due to $\mathcal{O}(1/N)$ corrections. For random free fermions we obtain:
\begin{equation}
\mathcal{C}_{ii}-f\,\simeq\, \mathcal{C}_{11}-f\,\,\,\,\,\longrightarrow \,\,\,\,\,t_{\textrm{random}}\sim \frac{1}{R}N^{\frac{1}{3}}\;. 
\end{equation}
Summarizing, from the present perspective there are three natural time scales one can consider: local relaxation time scales, information transport or scrambling, and near equilibrium randomization. For usual local systems, scrambling and randomization are equal, since the time that information takes to spread through the entire system is sufficiently big due to causality constraints. For non-local systems, such as random free fermions, the situation is different. Since information transport is instantaneous, local relaxation and scrambling are governed by the same time scale, while randomization takes a longer time.

In the next section we extend all this intuition to large-$\mathcal{N}$ matrix models.

\subsection{Information transport and scrambling with matrix models}

Let us recapitulate the important results obtained so far. In the large-$N$ limit of the random free fermion model, the correlator matrix is diagonal, see~(\ref{cort}) and~(\ref{diagonalCt}). The deviation from this diagonal behavior can be computed. This large-$N$ factorization happens even in time-dependent scenarios. It is rooted in the democratic structure of the microscopic Hamiltonian~(\ref{H}). Since the diagonal entries are given by the local expectation values of number operators $\langle c_{i}^{\dagger}c_{i}\rangle (t)=n_{i}(t)$, the entanglement entropy of a subset $A$ of $M\lesssim N/2$ degrees of freedom is simply given by:
\begin{equation}\label{sextensivefermions}
S_{A}^{\textrm{fermions}}=-\sum\limits_{i\in A}(\,n_{i}(t)\log n_{i}(t) + (1-n_{i}(t)\,)\log (\,1-n_{i}(t))\,)\;,
\end{equation}
For random particles, the entanglement entropy is an \emph{extensive} quantity even in time-dependent scenarios. The deviation from extensivity can be computed with the deviation of the correlator matrix from its diagonal mean. This computation shows that the deviation is subleading in the thermodynamic limit, and allows the study of entanglement in every subsystem, no matter its size, by the entanglement of on-site subsystems. As commented in the introduction, the generalization of the Hamiltonian~(\ref{H}) to the bosonic case is straightforward. The only change is the connection between entanglement entropy and occupation densities, the formula being slightly different. For random free bosons it would read:
\begin{equation}\label{sextensivebosons}
S_{A}^{\textrm{bosons}}=\sum\limits_{i\in A}(\,(n_{i}(t)+1)\log (\,n_{i}(t)+1)-n_{i}(t)\,\log n_{i} (t)\,)\;.
\end{equation}
We claim that the previous formulas extend to large-$\mathcal{N}$ matrix models and the SYK model \cite{subir}. The reason is that this formula is \emph{only} based on the effective gaussianity of reduced subsystems in democratic models. In other words, formulas~(\ref{sextensivefermions}) and~(\ref{sextensivebosons}) are information theoretic versions of large-N factorization and therefore apply to any theory satisfying such property (such as large-$\mathcal{N}$ matrix models and the SYK model). The only difference between the toy model and real microscopic descriptions of black holes lies in the specific time dependence of occupation densities. For random particles, we have a polynomial decay of number operators~(\ref{c11}) while for black holes we expect quasinormal ringing \cite{hubeny}.

Therefore, for the SYK model the extension is trivial. Since the Hilbert spaces are isomorphic, the nature of subsystems is the same. Entanglement evolution is just given by~(\ref{sextensivefermions}).

For large-$\mathcal{N}$ field theories one has to take care due to mainly two reasons. The first is that the democratic interactions appear only in color space. The second is that subsystems are not defined by partitions of the Hilbert space, since these partitions are not gauge invariant. Subsystems are better defined by operator algebras themselves. To avoid these two issues, the easiest path is to consider the theory in Fourier space. Our subsets will be sets of generalized free field modes\footnote{We refer to Ref.\cite{papadodimas} for more details and a complete set of references.} $O_{\omega,k}$. Generalized free fields are traces of products of the fields of the theory, in which we are not allowed to consider products with $\mathcal{O}(N)$ fields. Generalized free fields are gauge invariant operators by construction, due to the trace, a feature that avoids the problem of defining entanglement in a gauge invariant manner. They are called `generalized free fields' because they generate a Fock space of free particles in the large-$N$ limit, but they accomplish so without satisfying any free wave equation. This last feature is possible due to large-$N$ factorization \cite{thooft}. At finite temperature, for any set of generalized free fields $\mathcal{O}$, large-$N$ factorization is the following statement\footnote{We remark here that large-N factorization is rigorously proven in the vacuum \cite{thooft}. In what follows we assume its validity at finite temperature as well, as done in \cite{papadodimas}. This assumption is clearly supported by the AdS/CFT correspondence \cite{adscft}.}
\begin{equation}\label{thermalfact}
\textrm{Tr}(\, \rho_{\beta}\,\mathcal{O}(x_{1})\cdots \mathcal{O}(x_{2n})) = \frac{1}{2^{n}}\sum\limits_{\pi}\textrm{Tr}(\, \rho_{\beta}\,  \mathcal{O}(\pi_{1})\mathcal{O}(\pi_{2}))\cdots\textrm{Tr}( \,\rho_{\beta} \,\mathcal{O}(\pi_{2n-1})\mathcal{O}(\pi_{2n})) +\mathcal{O}(1/N)\;.
\end{equation}
As shown in \cite{papadodimas}, relation~(\ref{thermalfact}) allows us to define creation and annihilitation operators $O_{\omega,k}$ and $O_{\omega,k}^{\dagger}$ satisfying the usual algebra of thermal free oscillators. Assuming for concretenes a bosonic generalized free field we have:
\begin{equation}
C_{\omega k,\omega'k'}=Z^{-1}_{\beta}\,\textrm{Tr}(\, \rho_{\beta}\,O_{\omega',k'}^{\dagger}O_{\omega,k}\,)=\frac{1}{e^{\beta}-1}\,\delta (w-w')\,\delta (k-k')=\,n_{\mathcal{O}_{\omega,k}}^{\beta}\,\delta (w-w')\,\delta (k-k')\;,
\end{equation}
and
\begin{equation}
Z^{-1}_{\beta}\,\textrm{Tr}(\, \rho_{\beta}\,O_{\omega,k}O_{\omega',k'}^{\dagger}\,)=(1+n_{\mathcal{O}_{\omega,k}}^{\beta})\,\delta (w-w')\,\delta (k-k')\;,
\end{equation}
together with
\begin{equation}\label{crossterms}
Z^{-1}_{\beta}\,\textrm{Tr}(\, \rho_{\beta}\,O_{\omega',k'}O_{\omega,k}\,)=0 \,\,\,\,\,\,\,\,\,\,\,\,\,\,\, Z^{-1}_{\beta}\,\textrm{Tr}(\, \rho_{\beta}\,O_{\omega',k'}^{\dagger}O_{\omega,k}^{\dagger}\,)=0\;,
\end{equation}
where connected higher point correlation functions vanish. This implies that the two point correlation matrix $C_{\omega k,\omega'k'}$ is diagonal up to subleading corrections, sclaing as $\mathcal{O}(1/N)$. This whole discussion does not change if the quantum state $\rho$ is not the thermal density matrix, but an evolving pure/mixed $\rho (t)$ \footnote{In an AdS/CFT this is predicted from the gravitational perspective, since interactions in the bulk are suppressed in the large-$N$ limit}:
\begin{equation}
C_{\omega k,\omega'k'}(t)=\,\textrm{Tr}(\, \rho(t)\,O_{\omega',k'}^{\dagger}O_{\omega,k}\,)=\,n_{\mathcal{O}_{\omega,k}}(t)\,\delta (w-w')\,\delta (k-k')\;.
\end{equation}
So just by assuming~(\ref{thermalfact}), and repeating the argument of the previous section to bound the errors when computing entanglement entropy, the time dependent and gauge invariant entanglement entropy of \emph{any} set $A$ of $M\lesssim N/2$ generalized free fields is just given by:
\begin{equation}\label{SO2}
S^{A}(t)=\sum\limits_{i\in A}(n_{\mathcal{O}^{i}_{\omega,k}}(t) +1)\log (n_{\mathcal{O}^{i}_{\omega,k}}(t) +1)-n_{\mathcal{O}^{i}_{\omega,k}}(t) \log (n_{\mathcal{O}^{i}_{\omega,k}}(t))\pm \mathcal{O}(M^{2}/N)\;.
\end{equation}
Notice that due to large-$N$ factorization we are able to define entanglement in a gauge invariant manner. The clue is to consider entanglement entropy to be associated to observables or operator algebras \footnote{This might be of interest for the approach to entanglement entropy developed in \cite{amilcar}.}, such as generalized free fields $\mathcal{O}$. Given an operator algebra, there are many observations one can do (all possible correlation functions). Finding entanglement entropy from all correlator functions in the operator algebra can be quite complicated in general. Crucially, for large-$\mathcal{N}$ matrix models the situation simplifies due to large-$N$ factorization, and entanglement entropy can just be found by looking at the covariance matrix. This covariance matrix is diagonal in Fourier space and it is controlled only by occupation densities (one point functions), see~(\ref{SO2}).

At it should, formula~(\ref{SO2}) is coherent with the expectations for random pure states, where it correctly provides the thermal entropy at the appropriate temperature:
\begin{equation}
S^{\beta}_{A}=\sum\limits_{i\in A}\beta \omega_{i}\,\frac{e^{-\beta\omega_{i}}}{1-e^{-\beta\omega_{i}}}-\log (\,1-e^{-\beta\omega_{i}})\;.
\end{equation}
Besides, formula~(\ref{SO2}) implies that the conclusion obtained for random particles extends to matrix models:
\begin{itemize}
\item Entanglement entropies are fully controlled by the decay of occupation densities. Global scrambling of information is controlled by local relaxation of perturbations.
\end{itemize}
We want to stress again at this point that the only assumption we used to derive this result is large-$N$ factorization in thermal ensembles~(\ref{thermalfact}). This is well supported by the toy model of random particles and by the black hole geometric description of large-$N$ matrix models. Given this feature, reduced subsystems are products of uncorrelated gaussian density matrices. We can then use the covariance matrix approach for gaussian systems \cite{peschel}, since this is the only variable surviving the large-N limit. Finally, bounds on subleading terms can be found as in the previous section, from the size of the corrections to the two-point functions. 

Formula~(\ref{SO2}) is a remarkably simple result with many potential consequences. In what follows, we will just focus on its implications to information transport, scrambling, and randomization. To this end we first need to argue for the actual functional form of $n_{\mathcal{O}^{i}_{\omega,k}}(t)$ in large-$\mathcal{N}$ matrix models. Notice that on generic grounds, quasinormal ringing implies that the field mode decays at late times as:
\begin{equation}
\delta \mathcal{O}\sim \,e^{-\Omega_{\omega,k}^{\textbf{I}}t}\,(\,A'\,e^{i\,\Omega_{\omega,k}^{\textbf{R}}\,t}+B'\,e^{-i\,\Omega_{\omega,k}^{\textbf{R}}\,t})\;,
\end{equation}
where $\Omega_{\omega,k}^{\textbf{I}}$ and $\Omega_{\omega,k}^{\textbf{R}}$ are the real and imaginary parts of the quasinormal frequency $\Omega_{\omega,k}$. For a harmonic oscillator satisfying~(\ref{crossterms}), the number operator is basically the square of the field, so that we expect:
\begin{equation}\label{nmatrix}
n_{\mathcal{O}_{\omega,k}}(t)\simeq n_{\mathcal{O}_{\omega,k}}^{\beta}+\, A\,e^{-2\,\Omega_{\textrm{I}}\,t}\,(\,\cos (2\,\Omega_{\omega,k}^{\textbf{R}}\,t\,+\theta)\, +1\,)\;.
\end{equation}
Although the argument leading to~(\ref{nmatrix}) is heuristic, a similar law has indeed been found recently in \cite{occupation} for a related occupation density described first in the context of holography in \cite{vijayoccupation}. More importantly, we have recently verified that this type of law describes the evolution of Bekenstein-Hawking entropy \cite{aron}.

Plugging the previous expression~(\ref{nmatrix}) into~(\ref{SO2}), we obtain the evolution of entanglement entropy of a subset $A$ of generalized free fields. The formula is extensive, but each mode decays with a different $\Omega_{\omega,k}$. To analyze its properties we just need to analyze the evolution of single perturbations. Defining $f_{\omega,k}(t)\equiv \,A\,(\,\cos (2\,\Omega_{\omega,k}^{\textbf{R}}\,t\,+\theta)+1\,)$, the evolution of entanglement entropy associated to a single mode is given by:
\begin{eqnarray}\label{entmatrix}
S^{\mathcal{O}_{\omega,k}}(t)&=&(\,n_{\mathcal{O}_{\omega,k}}^{\beta}+\,e^{-2\,\Omega_{\textrm{I}}\,t}\,f_{\omega,k}(t) \,+1\,)\log (\,n_{\mathcal{O}_{\omega,k}}^{\beta}+\,e^{-2\,\Omega_{\textrm{I}}\,t}\,f_{\omega,k}(t) \,+1\,)-\nonumber\\ &-&(\,n_{\mathcal{O}_{\omega,k}}^{\beta}+\,e^{-2\,\Omega_{\textrm{I}}\,t}\,f_{\omega,k}(t)\,) \log (\,n_{\mathcal{O}_{\omega,k}}^{\beta}\,+e^{-2\,\Omega_{\textrm{I}}\,t}\,f_{\omega,k}(t)\,)\;.
\end{eqnarray}
We cannot trust the previous solution to small enough times since in that regime we should include higher quasinormal frequencies into~(\ref{nmatrix}). But notice that for $2\Omega_{\textrm{I}}t\ll 1$, the difference between the entanglement entropy at time $t$ and the one at time $t=0$ grows linearly with time:
\begin{equation}
S^{\mathcal{O}_{\omega,k}}\,(\,t\,\ll \,\frac{1}{2\,\Omega_{\textrm{I}}}\,)\propto \Omega_{\textrm{I}}\,t\;.
\end{equation}
On the other hand, for large times $2\,\Omega_{\textrm{I}}t\gg 1$, where formula~(\ref{nmatrix}) can be trusted, we enter the so-called entanglement plateaux, which is predicted to be:
\begin{eqnarray}\label{entmatrix1}
S^{\mathcal{O}_{\omega,k}}\,(\,t\gg \,\frac{1}{2\,\Omega_{\textrm{I}}}\,) &\simeq & S^{\beta}_{\mathcal{O}_{\omega,k}}+\,\log \,(\,\frac{n+1}{n}\,)\, e^{-2\,\Omega_{\textrm{I}}\,t}\,A\,(\cos (2\,\Omega_{\omega,k}^{\textbf{R}}\,t\,+\theta)+1\,)\,=\nonumber\\ &=& S^{\beta}_{\mathcal{O}_{\omega,k}}+ \,e^{-2\,\Omega_{\textrm{I}}\,t}\,\bar{A}\,(\cos (2\,\Omega_{\omega,k}^{\textbf{R}}\,t\,+\theta)+1\,)\;.
\end{eqnarray}
Relations~(\ref{entmatrix}) and~(\ref{entmatrix1}) are analogous to relations~(\ref{fermions}) and~(\ref{sc11}) for the case of random free fermions, and are part of the main results of the article. The difference between the toy model and the real one lies only in the functional form of the decay of occupation densities. A similar law was recently found to describe Bekesntein-Hawking entropy evolution, see Ref. \cite{aron}.

Finally, just for completeness, notice that the Mutual Information between gauge invariant generalized free field modes is again zero \emph{at all times}:
\begin{equation}
I\,(\,\mathcal{O}_{\omega,k}\, ,\mathcal{O}'_{\omega',k'}\,)\,(t)\sim\mathcal{O}(1/N)\:,
\end{equation}
due to the extensivity of entanglement entropies through unitary evolution. This is an information theoretic expression of large-N factorization. For the Mutual Information between bigger subsets of generalized free field modes one can use~(\ref{SO2}).

Joining all results together, the conclusions regarding scrambling and local relaxation are universal to all democratic models, including SYK and large-N field theories. The time scale associated to:
\begin{itemize}
\item Information lost by the generalized free field mode $\mathcal{O}_{\omega,k}$, quantified by its associated entanglement entropy,
\item Information spreading through the matrix model, characterized by the vanishing of connected correlators and Mutual Information between small subsystems at all times,
\item Relaxation of all entanglement entropies to thermal values at leading order,
\item Breaking of the mean field approximation, characterized by the growth of entanglement of one generalized free field mode.
\end{itemize}
is operator dependent, size independent and given by:
\begin{equation}
t_{\textrm{relax}}=t_{\textrm{scrambling}}=\frac{1}{2\,\Omega_{\textrm{I}}}\;,
\end{equation}
a result which put under question marks the fast scrambling conjecture \cite{susskind}. We want to stress again that although some aspects seem complicated, everything is based on large-N factorization, which implies factorization of reduced subsystems, and gaussianity.  Information theoretically, large-N factorization translates into equations~(\ref{sextensivebosons}),~(\ref{sextensivefermions}) and~(\ref{SO2}).

\subsubsection{Scrambling vs randomization time scales}

We have shown that random particles, SYK and large-$N$ matrix models behave very similarly in regards to information spreading. Entanglement entropy evolution is extensive for all of them, and it is fully controlled by the decay of occupation densities. This is rooted in large-N factorization. What differs in each model is the time dependence of the occupation densities themselves. Quite generically, we expect the occupation densities to decay exponentially fast, while for random particles we found only a polynomial decay\footnote{Notice that the polynomial decay might be connected with black holes in flat space, given the old results of Ref. \cite{flatold}. This is an exciting connection worth to explore.}. 

This difference can be seen through the randomization time. This was defined as the time scale by which deviations from thermality are uniform throughout the system. It can be found by looking at the properties of entanglement entropies at the entanglement plateaux, or equivalently by looking at the occupation densities. Physically, it is the time scale by which we cannot trust the time decay, since $\mathcal{O}(1/N)$ corrections kick in. For large-$N$ matrix models, it is just given by:
\begin{equation}
n_{\mathcal{O}_{\omega,k}}(t)-n_{\mathcal{O}_{\omega,k}}^{\beta}=\,e^{-2\,\Omega_{\textrm{I}}\,t}\,f_{\omega\, , k}\, (t)\sim \,\mathcal{O}(\frac{1}{N})\;,
\end{equation}
so that:
\begin{equation}\label{randommatrix}
t_{\textrm{random}}=\frac{1}{2\,\Omega_{\textrm{I}}}\log N \;.
\end{equation}
Taking into account the geometric observations done in \cite{us1,us3} concerning black hole geometric backgrounds, this randomization time scale is of the same order of magnitude as the flight crossing time through the near horizon region until the so-called brick wall \cite{brick}. Our results suggest that dynamical processes in the near horizon region correspond to near equilibrium evolution, fully controlled by quasinormal frequencies, and cut \emph{dynamically} at the time scale~(\ref{randommatrix}). The remarkable point is that the saturation of the process at such randomization time scale is not a phenomenological assumption, unlike in the gravity side. In the present perspective, the saturation at~(\ref{randommatrix}) is a purely \emph{dynamical} feature, due to finite $N$ effects coming from unitarity of the microscopic model. Notice also that we expect a different randomization time for each perturbation. The emergence of the brick wall might be not universal after all.  In this framework, it definitely depends on the specific observable we use to probe the black hole.

\subsection{Comments on unitarity breaking in the thermodynamic limit}

In this section we comment on another interesting and distinguishing feature of democratic systems. Going to~(\ref{cort}), and taking the large-$N$ limit we obtain:
\begin{equation}
\mathcal{C}_{ij}(t)^{\infty}\equiv\lim_{N\to\infty}\mathcal{C}_{ij}(t)=f\delta_{ij}+(n-f)\,\frac{4\,(J_{1}(Rt))^{2}}{(Rt)^{2}}\,\delta_{i1}\,\delta_{j1}\;.
\end{equation}
Since the resulting correlation matrix is diagonal, the Von Neumann entropy associated to it is given by:
\begin{equation}
S(\rho)=S_{1}(t)+S_{BH}=S(t)\;,
\end{equation}
a time-dependent function. In the large-$N$ limit, the entropy associated with the \emph{global} state changes with time. Its evolution cannot be driven by a unitary operator, and should be represented by some super-evolution operator $\$ $, in the lines of \cite{paradox}. In this precise sense, unitarity is \emph{globally} broken in the thermodynamic limit. We stress that this is essentially different from local theories, where superoperators and non-unitary behavior appear only \emph{locally} when tracing out some part of the system. Taking the thermodynamic limit $N\to\infty$ does not affect the entropy globally as time evolves in local systems.

We conclude that there is in principle no contradiction whatsoever between unitarity of quantum gravity and non-unitarity of the Einstein-Hilbert action (its semiclassical limit), whenever the microscopic theory is democratic, such as Matrix Models \cite{matrix}, AdS/CFT \cite{adscft}, or the recently introduced infinite range fermionic model \cite{subir}.

\section{Conclusions}\label{secIV}

In this article we first have proposed and studied a quantum toy model of black hole physics. While capturing essential features of quantum black holes, the model is still exactly solvable. The two properties we wanted our system to possess are quantum thermalization and democracy of interactions. As described in the introduction, `democracy of interactions' means the strongest form of non-locality, which amounts to have every oscillator interacting with every other oscillator through couplings of the same size. The first requirement is an obvious necessary feature since there are very strong reasons to consider black holes as many body systems at finite temperature \cite{hawking,witten}. The second property intends to emulate realistic black hole models, such as large-$\mathcal{N}$ matrix models \cite{matrix,adscft,susskind,subir}. The color sector physics of large-$\mathcal{N}$ matrix models furnish an example of such democratic systems.

The proposed toy model was defined by the Hamiltonian~(\ref{H}). It is a gaussian system of coupled fermionic oscillators in which the mass matrix belongs to the Gaussian Orthogonal Ensemble of random matrices \footnote{See  \cite{tao,haake} for excellent and complete treatments of Random Matrix Ensembles from a mathematical and physical perspective respectively. See the third footnote of the article for comments on modifying the nature of the randomness of the model.}. This system was described in \cite{useth}, as an analytical example of the Eigenstate Thermalization Hypothesis \cite{srednickisub,ETH,eisertreview}. The advantages of choosing such a model are the following:
\begin{itemize}
\item The model is exactly solvable. Corrections of $\mathcal{O}(1/N)$ can be properly taken into account.
\item The model has enough complexity to display quantum thermalization. It allows a precise study of information spreading in democratic systems, at the global and local levels.
\item The extension to bosons is straightforward.
\end{itemize}
Within the toy model, we have chosen to analyze two generic physical processes. The first scenario is the usual black hole thought experiment, in which one throws in a particle initially unentangled with the black hole~(\ref{initial}). Subsequent evolution entangles both particle and hole, and information about the initial state $n$ is mixed/spread through the system (but conserved due to unitarity). This unitary time evolution can be fully accounted by the two-point correlation functions of the theory,  equations~(\ref{S}),~(\ref{dS}),(\ref{cort}), and~(\ref{entt}). The second scenario is akin to a `collapse' scenario, in which the initial state is a factorized many-particle state~(\ref{factorized}). For this case, the computations are harder, but we could show that the typical correlation matrix is again diagonal~(\ref{diagonalC}). 

As described in the second part of the article, this type of covariance matrix appears also in realistic models of black holes, such as SYK or large-$N$ matrix models. The property that underlies this covariance matrix is large-N factorization, a property characteristic of both SYK and large-$N$ matrix models that seems to be a defining feature of democratic models.

Given large-N factorization, extensivity of entanglement evolution follows naturally, with subleading computable corrections to it. The formulas for all the models are equivalent, and given by~(\ref{sextensivefermions}) for random free fermions or the SYK model, by~(\ref{sextensivebosons}) in the random bosonic cases and by formula~(\ref{SO2}) for matrix models, where it is more convenient to work with generalized free field modes in Fourier space. An intuitive explanation of the extensivity of entanglement evolution is the following. Due to the non-locality of the theory, correlations between a given degree of freedom and any other are all of the same size. Since there are $\mathcal{O}(N)$ degrees of freedom interacting at the same time and given monogamy of entanglement, the shared correlation between any pair of them must be of $\mathcal{O}(1/N)$. Large-$N$ factorization seems thus to be related with monogamy of entanglement in democratic systems.

In every case, the main conclusions are:
\begin{itemize}
\item In the thermodynamic limit, the behavior of entanglement entropies is extensive at all times, see~(\ref{sextensivebosons}) and~(\ref{sextensivefermions}). The evolution of entanglement entropies is fully controlled by local relaxation of occupation numbers.
\item Information spreading is instantaneous. Any lost information is instantaneously scrambled since the Mutual Information between small enough subsystems is zero at all times.
\item The time scale by which the entanglement entropy of a subset $A$ reach the entanglement plateaux is independent of the system size and it is controlled by on-site quantities. The size $A$ can scale with the size of the total system, as long as it is not bigger than half of it.
\item The mean field approximation, defined as the time of on-site entanglement production is broken at a time independent of the system size, although another kind of mean field approximation is valid for longer times~(\ref{mmf}).
\end{itemize}
The previous results challenge the fast scrambling conjecture \cite{susskind}. As described in section~(\ref{secIII}), this conjecture states that there is a minimum lower bound for the time scale associated to `information scrambling'. The conjectured lower bound is of order $\mathcal{O}(\beta\log N)$. Information scrambling was defined by global equilibration of entanglement entropies, together with information spreading through the entire system. The previous results suggest that both requirements are fulfilled in a time $t_{\textrm{scrambling}}=\frac{1}{2\,\Omega_{\textrm{I}}}$, where $\Omega_{\textrm{I}}$ is the lowest quasinormal frequency associated with the subset $A$ that has been brought out of equilibrium. Although there is still room for discussion of the present results, and probably subtle aspects concerning information theory might appear given that we are dealing with very unconventional Hamiltonians, the present results seem to be making a strong case against the fast scrambling conjecture and they invalidate the mean field bounds presented in \cite{lashkari}.

We conclude that the only difference between the different models lies in the specific time dependence of number operators. Whereas for random particles we found polynomial decays, for matrix models we have exponential decays. Therefore, the time scale by which $\mathcal{O}(1/N)$ corrections need to be taken into account is different in each case. This time scale was coined `randomization time'. It is the time scale by which the deviations from thermality are uniform over the system. For random particles, we found a randomization time scaling polynomially with $N$. For matrix models, the result turns out to be:
\begin{equation}
t_{\textrm{random}}=\frac{1}{2\,\Omega_{\textrm{I}}}\log N\;.
\end{equation}
Interestingly, this time scale sets the causality bounds for crossing the near horizon region until the so-called brick wall or stretched horizon, see \cite{us1,us3}. It is tempting to conclude that the emergence of the near horizon geometry is related to near equilibrium evolution, characterized by the properties of the entanglement plateaux, or analogously by the properties of the occupation densities. These two quantities are controlled by the famous quasinormal modes, so it is tempting to conclude that the emergence of near horizon regions is fully encoded in the quasinormal ringing of the dual field theory, i.e in the poles of retarded correlation functions \cite{hubeny,ivosachs}. This speculative proposal provides a \emph{dynamical} reason for the appearance of the brick wall, as the time by which the evolution of occupation densities and entanglement entropies is affected by $\mathcal{O}(1/N)$ corrections.

In the last section, we made a remark about the information paradox \cite{paradox} in the light of the toy model. We noticed that unitarity of the toy model, characterized for example by a time independent Von Neumann entropy of the global state, is lost in the large-$N$ limit \emph{at all times}. Taking the large-N limit affects the entropy globally. This is a characteristic feature of democratic systems, whose semiclassical approximation at high energies is effectively described by non-unitary dynamics. The old and famous proposal of \cite{paradox} seems to naturally emerge in theories of quantum gravity which have a democratic structure of interactions.

\section*{Acknowledgements}

It is a pleasure to thank Aron Jansen and Stefan Vandoren for interesting discussions.
This work was supported by the Delta-Institute for Theoretical Physics (D-ITP) that is funded by the Dutch Ministry of Education, Culture and Science (OCW).

%{toc}{chapter}
%\addcontentsline{toc}{section}{Bibliography}

%bibliography{bibmutual} % texto.bib es el fichero donde est� salvada la bibliograf�a.
%\bibliographystyle{unsrt} % estilo de la bibliograf�a.

%\addcontentsline{toc}{section}{Bibliography}

%\bibliography{bibuscodif} % texto.bib es el fichero donde está salvada la bibliografía.
%\bibliographystyle{unsrt} % estilo de la bibliografía.
\end{document}